%% file: Pollock.Mk34.T-ReX.arXiv.tex
\documentclass[fleqn,usenatbib]{mnras}

\usepackage[T1]{fontenc}
\usepackage{ae,aecompl}

\usepackage{graphicx}	
\usepackage{amsmath}	
\usepackage{amssymb}	

\hypersetup{pdfauthor={A. M. T. Pollock},
               pdftitle={The 155-day X-ray cycle of the very massive Wolf-Rayet star Melnick 34 in the Large Magellanic Cloud},
               pdfkeywords={stars: Wolf-Rayet, stars: massive, stars: winds, outflows, binaries: eclipsing, X-rays: stars, shock waves},
               bookmarksnumbered=true}

\newcommand{\Chandra}{\textit{Chandra}}
\newcommand{\Swift}{\textit{Swift}}
\newcommand{\XMM}{\textit{XMM-Newton}}

\newcommand{\mr}[1]{\mathrm{#1}}
\newcommand{\NaN}[1]{\multicolumn{#1}{c}{...}}

\newcommand{\changed}[1]{\textnormal{#1}}

\title[155-day X-ray cycle of Mk~34]{The 155-day X-ray cycle of the very massive Wolf-Rayet star Melnick 34 in the Large Magellanic Cloud}

\author[A. M. T. Pollock et al.]{
A. M. T. Pollock,$^{1}$\thanks{E-mail: A.M.Pollock@sheffield.ac.uk}
P. A. Crowther,$^{1}$
K. Tehrani,$^{1}$
\newauthor
Patrick S. Broos,$^{2}$
and
Leisa K. Townsley,$^{2}$
\\
$^{1}$Department of Physics and Astronomy, University of Sheffield, Hounsfield Road, Sheffield S3 7RH, England\\
$^{2}$Department of Astronomy \& Astrophysics, 525 Davey Laboratory, Pennsylvania State University, University Park, PA 16802, USA
}

\date{Accepted XXX. Received YYY; in original form ZZZ}

\pubyear{2018}
\volume{{\rm in press}}

\begin{document}
\label{firstpage}
\pagerange{\pageref{firstpage}--\pageref{lastpage}}
\maketitle

\begin{abstract}
The Wolf-Rayet star Mk~34 was observed more than 50 times
as part of the deep T-ReX \Chandra~ACIS-I X-ray imaging survey of the Tarantula Nebula in the Large Magellanic Cloud
conducted between 2014 May and 2016 January.
Its brightness showed one bright maximum and repeated faint minima which help define an X-ray recurrence time of $155.1\pm0.1$ days
that is probably the orbital period of an eccentric binary system.
The maximum immediately precedes the minimum in the folded X-ray light curve as confirmed by
new \Swift~XRT observations.
Notwithstanding its extreme median luminosity of $1.2\times10^{35}\mr{erg}~\mr{s}^{-1}$, which makes it over an order of magnitude
brighter than comparable stars in the Milky Way,
Mk~34 is almost certainly a colliding-wind binary system.
Its spectrum shows phase-related changes of luminosity and absorption that are probably related
to the orbital dynamics of two of the most massive stars known.
\end{abstract}




\section{Introduction}


The Tarantula Nebula or 30 Doradus in the Large Magellanic Cloud is the most important star-forming complex in the Local Group.
At its heart is R136, the stellar cluster that contains the most massive stars known \citep{CSH+:2010,CCB+:2016}
which fall into the categories of O stars and more particularly hydrogen-rich Wolf-Rayet stars.
Between 2014 and 2016, a high-spatial-resolution X-ray imaging survey known as T-ReX
was undertaken with the \Chandra~Observatory
to study both the stellar population itself and the thermal and dynamical effects
wrought by stellar winds and supernova shocks on the surrounding interstellar medium.

The exposure time of 2 million seconds accumulated over nearly 21 months offers a far more intensive view of the stars in 30 Doradus than
anything so far available for the Milky Way although there have been significant amounts of X-ray observing time on individual massive
stars such as the nearby, single, early O-type supergiant
$\zeta$~Puppis \citep[e.g.][]{NFR:2012},
long used by \XMM~as a calibration source,
and the more distant long-period binaries
WR~140 \citep[e.g.][]{P:2012}
and $\eta$~Carinae \citep[e.g.][]{C:2005}. 
There is a clear physical distinction between the soft intrinsic X-ray emission produced in the winds of single stars,
commonly attributed to many microscopic shocks supposed by \citet{FPP:1997} and others to form as a consequence of instabilities in the wind-driving mechanism,
and the harder, more luminous emission resulting from the macroscopic shock interaction between the counterflowing winds in a bound system of two or
more massive stars \citep[e.g.][]{SBP:1992, RN:2016}.

The variability properties of single and binary stars are also different. The \XMM~data of $\zeta$~Puppis collected intermittently
over many years have shown stochastic variability of unknown origin of up to 10\% or so on timescales of hours or days \citep{NOG:2013}
that appears typical of the intrinsic emission of single O stars in general.
Among the binaries, on the other hand, the highly eccentric system WR~140, of 7.9-year orbital period, shows changes in X-ray intensity many times
higher determined mainly by the relative disposition of the two stars through orbital separation and stellar conjunction \citep[e.g.][]{P:2012,SMT+:2015}.
Closer binary systems with periods, $P$,
of days rather than years,
such as the Wolf-Rayet binary
V444~Cygni (WN5o+O6III-V, $P=4.2\mr{d}$), can show high-amplitude phase-repeatable X-ray variability, more complex than at longer wavelengths,
where the hardness of the spectrum suggests colliding winds are involved \citep{JNH+:2015}.
On the other hand, X-ray emission from the prominent visual eclipsing O-star binary $\delta$~Orionis ($P=5.7\mr{d}$)
looked more like a single star in its behaviour during a simultaneous X-ray and optical campaign \citep{NHC+:2015} that covered most of an orbital cycle.

\begin{figure}
\includegraphics[width=\columnwidth,viewport=130 40 670 525,clip]{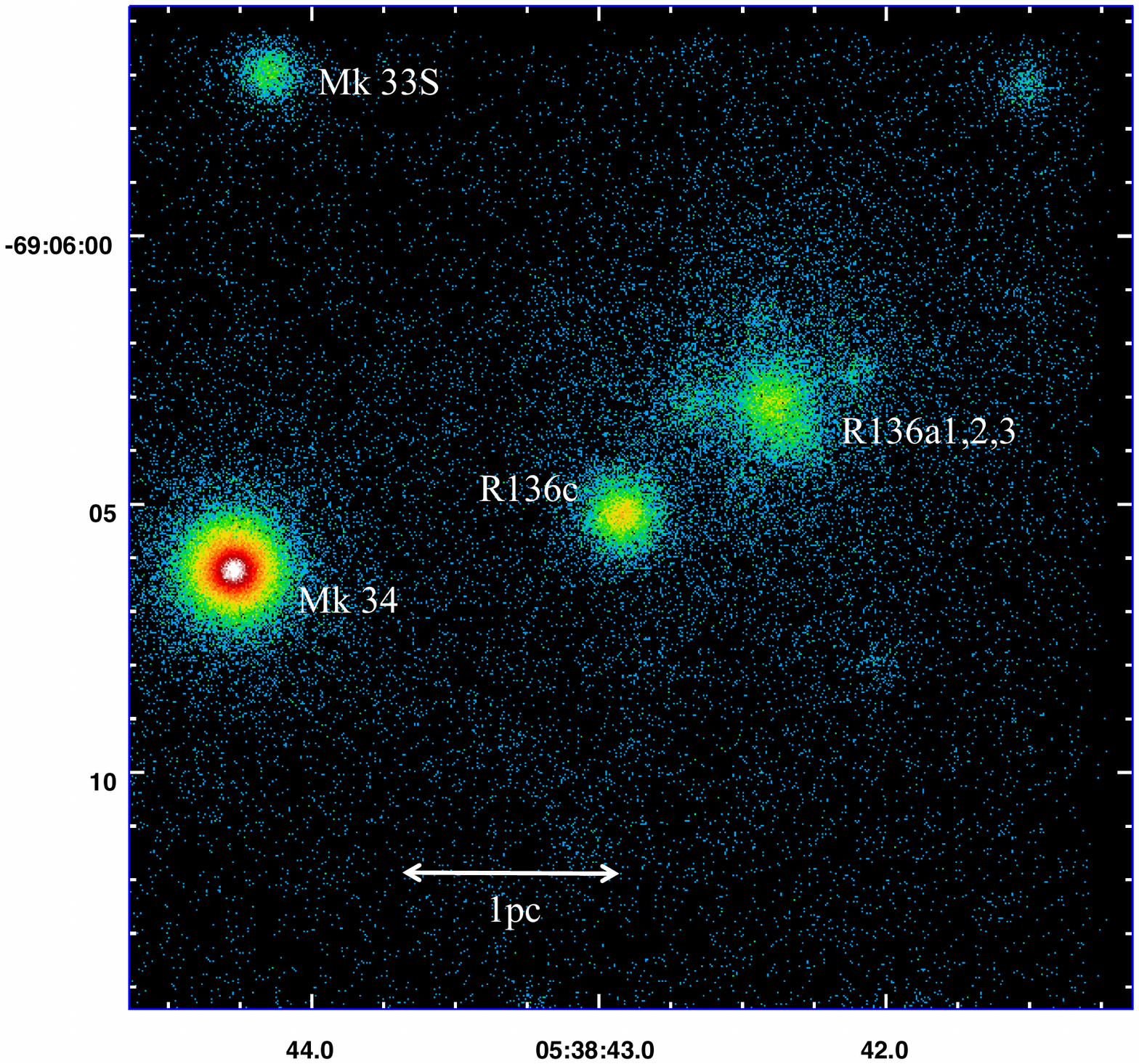}
\caption{\Chandra~ACIS-I X-ray image on a logarithmic intensity scale of the core of 30 Doradus accumulated
              during the T-ReX campaign of an area of 19\arcsec$\times$19\arcsec centred on R136c. North is up and east to the left.
              Also shown is the linear length scale at the 50~kpc distance of the LMC.}
\label{Fig:R136}
\end{figure}

For single and binary stars,
variability in any part of the electromagnetic spectrum is an observational property of significance for determining the physical nature of a system.
The T-ReX survey,
which will be described in detail elsewhere,
offers unprecedented opportunities of this type in the X-ray band.
This paper concentrates on one
particular star, Melnick 34 or BAT99 116, abbreviated Mk~34,
near the centre of the broader survey.
It lies 10\arcsec or 2.5~pc projected distance
east of R136, the dense stellar core of 30 Doradus from which,
as shown in Figure~\ref{Fig:R136}, it is comfortably resolved by the \Chandra~telescope with the ACIS-I instrument 
which offers by far the best X-ray resolving power available now and in the foreseeable future.

Mk~34 is a hydrogen-rich Wolf-Rayet star, classified WN5ha by \cite{CW:2011} or WN5h:a by \cite{HRT+:2014}, who assigned a high mass of $390~\mathrm{M}_{\sun}$
through spectroscopic analysis.
In the VLT-FLAMES census of the hot luminous stars in 30 Doradus \citep{DCdK+:2013},
similar analysis suggested Mk~34 to be the second most luminous star
behind R136a1 and to have the highest mass-loss rate
\changed{of $6.3\times10^{-5}\mr{M}_{\sun}/\mr{yr}$}
among 500 confirmed hot stars\changed{, about 20\% higher than the estimate of \cite{HRT+:2014}}.
The colour index $B-V=0.25$ shows that the star suffers significant extinction and X-ray absorption in the interstellar medium.

Through the discovery of dramatic optical radial-velocity variations that increased smoothly over 11 days only to have vanished a week later,
\citet{CSC+:2011} established Mk~34, also designated P93\_1134, to be a highly eccentric binary of two very massive stars.
The data plotted in the middle panel of their Figure 1 suggest an orbital period
greater than 50 days and a mass ratio of about $0.8$.

Previous X-ray measurements showed it was the brightest stellar X-ray source in 30 Doradus, designated ACIS~132 of the 180 sources in the complete
1999 \Chandra~study
of \cite{TBF+:2006} and CX~5 of the 20 bright objects discussed by \cite{PZPL:2002}.
Before any published optical radial-velocity evidence, their common speculation was that Mk~34 is a binary with X-rays powered by colliding winds,
this despite a luminosity about an order of magnitude higher than any comparable system in the Milky Way
and no
evidence for variability, either in the \Chandra~data or in comparison with other measurements years earlier.

\section{\Chandra~T-ReX Campaign}

The 51 observations of the \Chandra~Visionary Program (PI: Townsley)
known as T-ReX, to signify the Tarantula -- Revealed by X-rays,
were executed over 630 days between 2014 May 3 and 2016 January 22.
They were all targeted at R136a1, the central star of the dense stellar core of 30~Doradus that lies 10\arcsec~from Mk~34,
and were executed at different
spacecraft roll angles according to season.
The ACIS-I field-of-view measures 16.9\arcmin$\times$16.9\arcmin~so that Mk~34 was always located in the central parts
of the detector where the angular resolution is at its sharpest.
The disposition of sources including Mk~34 with respect to detector edges or bad-surface geometry depends on the roll angle which
thus partly determines detection sensitivity.
Each observation was analysed with the ACIS Extract package \citep{BTF+:2010}
used in previous work on stellar clusters,
notably the \Chandra~Carina Complex Project \citep{TBC+:2011,BTF+:2011}.
For point sources such as Mk~34,
the package delivers overall X-ray count rates and medium-resolution spectra
with accompanying calibration material dependent on individual observing conditions
to allow robust comparison of measurements taken at different times.
For Mk~34 in this paper, the detected count-rates were corrected according to
the relative mean of the energy-dependent effective area tabulated for each
observation in the ARF file, the so-called associated response function, that also encodes the instrumental geometry.

The observation log in Table~\ref{Table:ChandraLog} shows the sequence of sensitivity-corrected count rates for Mk~34 during the T-ReX
campaign with
three earlier archived observations from 2006 January
bringing the total to 54. 

As discussed by \citet{BTF+:2011} in their Appendix A and \citet{TBG+:2014}, sources observed with the \Chandra~ACIS instrument are subject to pile-up
in which two or more photons detected in identical or adjacent spatial and temporal readout elements are indistinguishable from single events
with the energies combined. These are then either accepted or discarded according to geometrical criteria known as grade selection.
The outcome of these two effects is a reduction in the count rate
that scales linearly with source brightness in Mk~34
and spurious hardening of the spectrum.
For each X-ray source, the ACIS Extract package provides simulated countermeasures in an overall
count-rate correction factor and a spectrum restored to remove pile-up distortion.
Mk~34 has been bright enough during most of its \Chandra~observational history in Table~\ref{Table:ChandraLog} for pile-up to be significant:
the count-rate correction factor increases linearly with detected count rate to a maximum of 1.17 about a median of 1.08.
While it is important to be aware of the extent of pile-up, the temporal analysis
discussed below is most reliably done without pile-up corrections which have therefore not been applied to
the count rates in Table~\ref{Table:ChandraLog}.
On the other hand, use of
the reconstructed spectra is unavoidable for the X-ray spectroscopy considered below in section~\ref{section:spectroscopy}.

\input{Chandra.log}

\subsection{\Chandra~X-ray photometry of Mk~34}\label{section:photometry}

\begin{figure}
\includegraphics[width=\columnwidth]{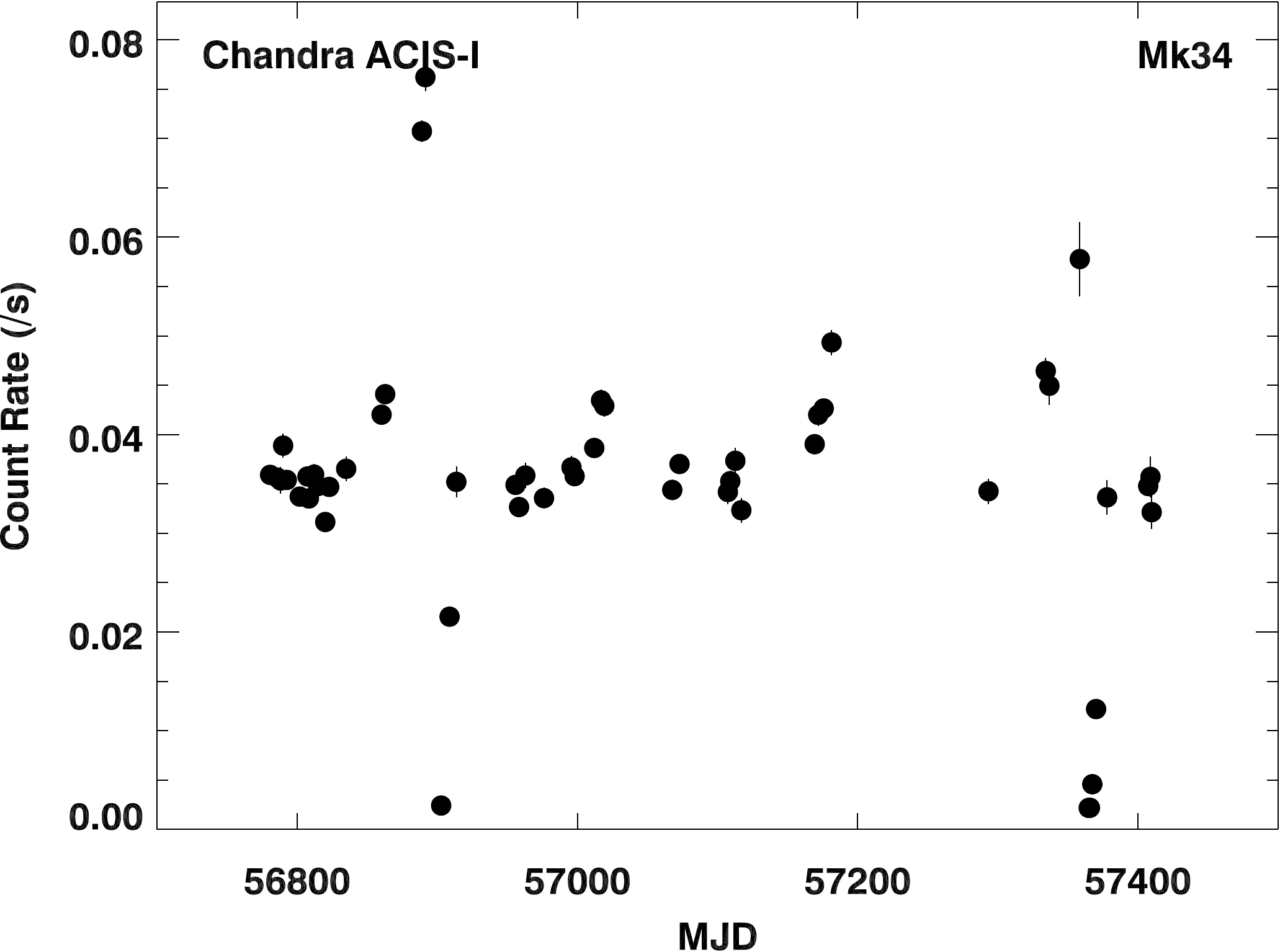}
\caption{\Chandra~ACIS-I X-ray count rates of Mk~34 during the T-ReX campaign. \changed{Many of the error bars are smaller than the plot symbol.}}
\label{Fig:Photometry}
\end{figure}

The 54 observations of Mk~34 had exposure times between 9.8 and 93.7~ks about a median of 37.6
with count rates ranging between 2.2 and 76.2 counts per ks about a median of 35.8.
The 51 count rates of Mk~34 during the 630-day T-ReX campaign itself are plotted in Fig.~\ref{Fig:Photometry} and show
obvious variability. This first became clear when two high measurements separated by 3 days in 2014 August at about twice the
median
were followed 11 days later by a measurement close to zero and a further 6 days later at about half the median.
A second minimum stretching over about a week in 2015 December and other brighter measurements in Fig.~\ref{Fig:Photometry} strongly suggest
4 cycles of repeatable structure in the course of the 630-day T-ReX campaign.

\begin{figure}
\includegraphics[width=\columnwidth]{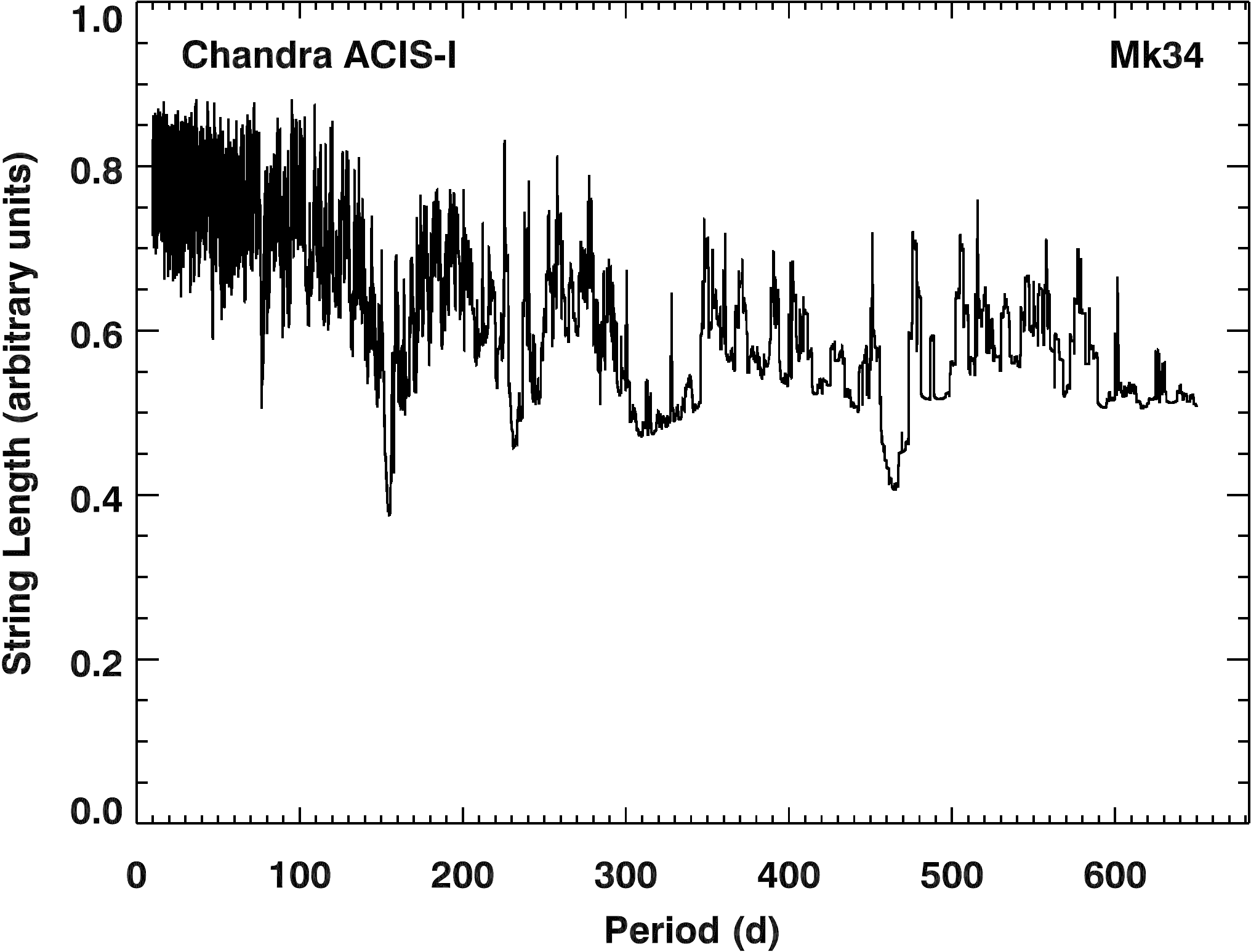}
\caption{String length against trial period for the 54 \Chandra~ACIS-I X-ray count rate measurements of Mk~34 in Table~\ref{Table:ChandraLog}.}
\label{Fig:StringLength}
\end{figure}

Given the irregular sampling of the light curve, a more precise value was explored through minimum string-length analysis
of the quantities
$$\{x_i,y_i\}=\{r_i\cos{\phi_i},r_i\sin{\phi_i}\}$$
where $r_i$ is the count rate of the $i$th measurement ordered in folded phase of a trial period.
The results are plotted in Fig.~\ref{Fig:StringLength}.
The minimum value of the string length curve infers a best-value period of 155.1 days with an uncertainty of 1 or 2 days.
We also considered the Plavchan algorithm \citep{PJK+:2008} implemented by the
NASA Exoplanet Archive\footnote{http://exoplanetarchive.ipac.caltech.edu}.
Although this method is considered by its authors better suited to the detection of small-amplitude variations among much more
numerous data than available for Mk~34,
it gave similar results.

\begin{figure}
\includegraphics[width=\columnwidth]{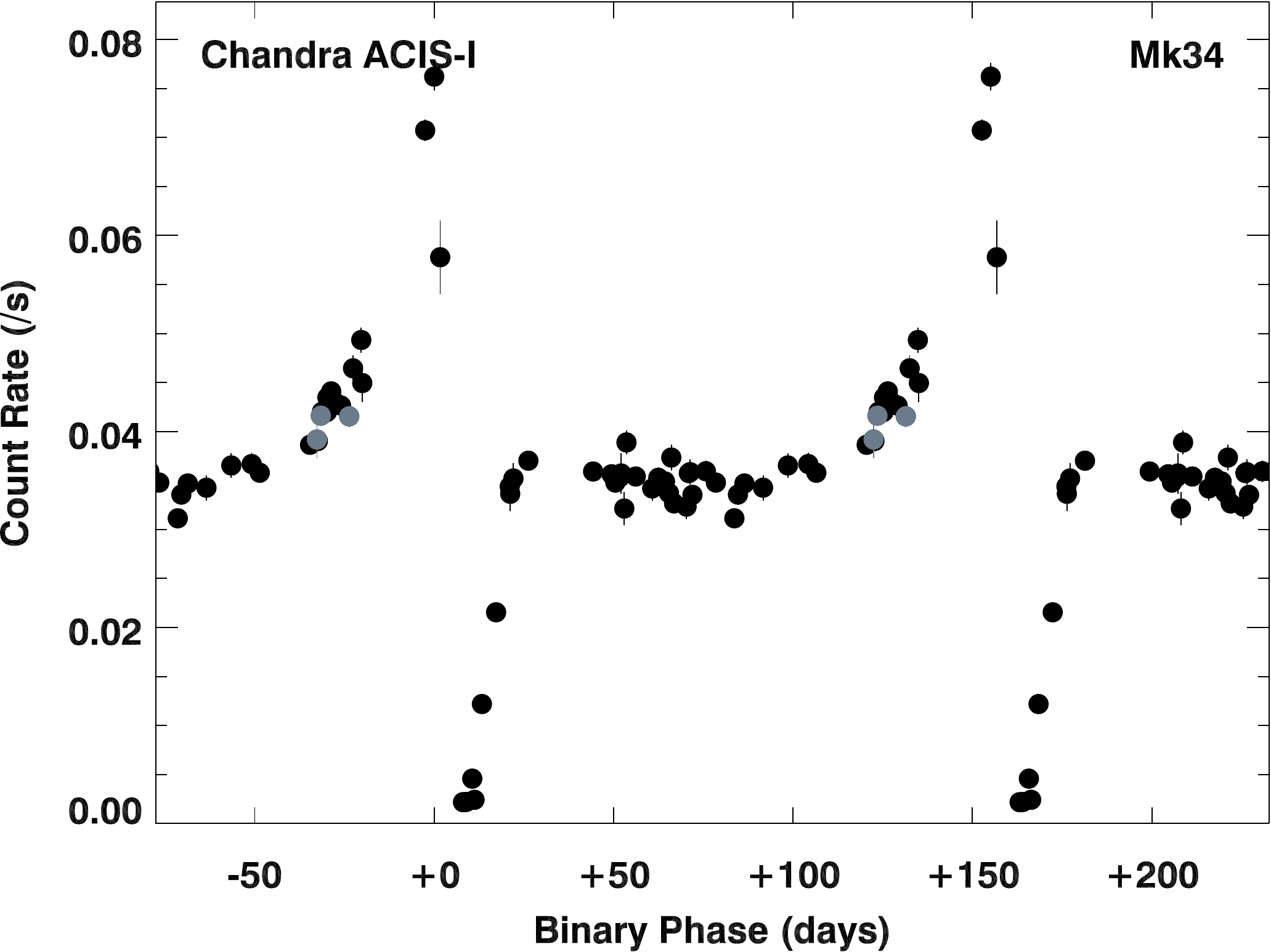}
\caption{Repeated cycles of \Chandra~ACIS-I X-ray count rates of Mk~34 folded at the period of 155.1 days derived from string-length analysis
              centred on X-ray maximum. The points plotted in grey are those from the 3 archived observations in 2006 January, 8 or more years before
              the other data.}
\label{Fig:X-rayCycle}
\end{figure}

The folded light curve with $P=155.1\rm{d}$ is shown in Fig.~\ref{Fig:X-rayCycle}, where the maximum count rate was taken to define zero phase.
The folded light curve shows accurate repeatability:
despite some missing coverage, the light curve clearly shows an accelerating 30 or 40-day rise to maximum, followed by a sharp
decrease in a few days to a minimum that lasts a few days, before a steady 10-day recovery to a relatively stable state of probably more than
100 days in length. Count rates plotted in grey from the archived observations over 10 days in late 2006 January
agree within small errors
with those close in phase taken 8-10 years later as the rise began to accelerate.
The decrease after maximum appears very steep: in the folded light curve the closest measurements were $1.6$ and $7.9$ days later
when the count rate had fallen by factors of $1.3$ and more than $30$, respectively.

There is little evidence for rapid variability within individual observations as shown by the modest values of $\log{\mr{P}_\mr{KS}}$
reported in Table~\ref{Table:ChandraLog}, where $\mr{P}_\mr{KS}$ is the {\em P}-value for the one-sample Kolmogorov-Smirnov statistic
under the hypothesis of constant source flux.

\subsection{X-ray historical record}

The \XMM~science archive\footnote{http://nxsa.esac.esa.int/nxsa-web/\#search} shows that Mk~34 has been detected 3 times by the EPIC imaging spectrometers
with the count rates reported in Table~\ref{Table:XMMLog}
during observations of PSR~J0537-6909 in 2001 November, IGR~J05414-6858 in 2011 October and in one exposure in 2012 October of the 48 forming
a wide-area survey of the LMC.
The three EPIC instruments, pn, MOS1 and MO2, observe simultaneously and cover nearly identical fields-of-view although
Mk~34 was not observed by the pn instrument in 2001 November because of the timing mode chosen for the pulsar
or by the MOS instruments in 2011 October because the star fell very close to the edge of the pn field-of-view where MOS coverage does not reach.
Despite these complications, the XMM measurements of Mk~34, whose identification is confused as 3XMM~J053843.[79]-690605 in the 3XMM-DR6 catalogue,
showed clear variability by about a factor of 2 very well correlated with the
\Chandra~folded light curve as shown by comparison with the ACIS-I count rates closest in phase also reported in Table~\ref{Table:XMMLog}.
The 3 observations all took place in the rising part of the X-ray light curve with the 2001 measurement, the brightest of the three, a few days
before maximum light 30 cycles earlier than the T-ReX maximum
suggesting an uncertainty in the period of the order of 0.1 days.

\input{XMM.log}

\subsection{New \Swift~X-ray photometry}

Once the T-ReX campaign had finished, Mk~34's putative period of 155.1 days was used to predict the timing of the next maximum in early 2016 May.
With the recognition that the source is strong enough for sufficiently accurate measurements in reasonable exposures with the \Swift~XRT instrument in
imaging PC mode,
an application for \Swift~ToO observations was submitted and approved to cover the anticipated maximum and subsequent minimum at intervals of about 7 days.
The results are shown in Table~\ref{Table:SwiftLog} where the sequence of detected count rates confirms the expectations based on the
\Chandra~folded light curve in Figure~\ref{Fig:X-rayCycle}: the highest count rate detected on 2016 May 3 within hours of the predicted maximum was about twice
as bright as the measurement about a month later. The much lower intervening count rates did not match the reductions of factors of 30 or so observed with
\Chandra, probably because of source confusion within the more modest angular resolution of the XRT which, in common with any other current X-ray instrument,
does not match that of ACIS images. \changed{The \Swift~XRT images show two clearly resolved sources coincident with Mk~34 and R140a that are
separated by 54\arcsec~but cannot distinguish between Mk~34 and 
R136c only 7\arcsec~away but a median factor of 8.6 fainter in resolved \Chandra~images.}

\input{Swift.log}

\section{\Chandra~X-ray spectroscopy of Mk~34}\label{section:spectroscopy}

\begin{figure}
\includegraphics[width=\columnwidth]{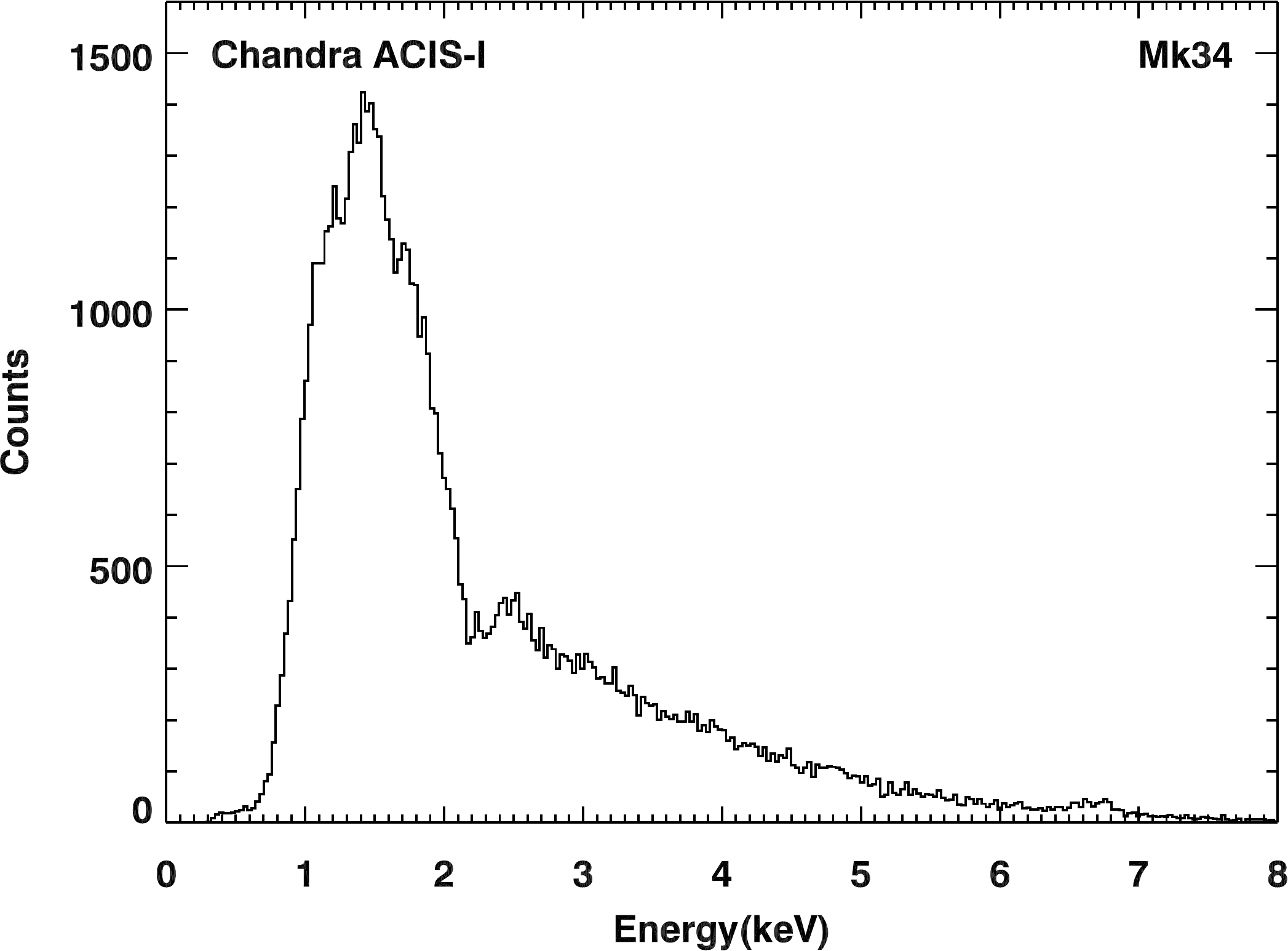}
\caption{\Chandra~ACIS-I X-ray spectrum of Mk~34 accumulated from all the available data.}
\label{Fig:Spectrum}
\end{figure}

The accumulated moderate resolution ACIS-I X-ray spectrum of Mk~34 shown in Figure~\ref{Fig:Spectrum} gives a clear qualitative
picture of its general spectral properties. The spectrum is hard,
stretching beyond the clear detection of \ion{Fe}{XXV} at 6.7 keV with the presence of a strong continuum in addition to
emission lines at all energies.
This type of spectrum is characteristic of well-established colliding-wind binaries such as
WR~140 \citep[e.g.][]{PCSW:2005},
WR~25 \citep[e.g.][]{PC:2006}
and $\eta$~Carinae although, given the 50 kpc distance of the LMC, Mk~34 is more luminous.
In comparison with WR~25 at about 2.3 kpc in Carina, for example, Mk~34 is more than 20 times further away and normally only 10 times fainter,
suggesting a luminosity greater by roughly an order of magnitude.
The much weaker intrinsic emission of some single Wolf-Rayet stars such as
WR~78 \citep{SZG+:2010}
and WR~134 \citep{SZG+:2012}
also show He-like line emission of \ion{Fe}{XXV} from very hot plasma but with no obvious continuum
which therefore appears to be an important defining property of colliding winds.
At low energies, the spectrum of Mk~34 is marked by a steep cutoff below about 1 keV,
likely due to a combination of circumstellar and interstellar photoelectric absorption,
leaving little
flux at the energies most affected by the instrumental contamination discussed below.

Each of the 54 individual spectra that contribute to Figure~\ref{Fig:Spectrum} is accompanied by customised response files
that reflect the
contemporary calibration of the ACIS instrument
and enable detailed models of the observed spectra to be constructed. Of particular relevance is knowledge of the
instrumental contamination\footnote{http://cxc.harvard.edu/ciao/why/acisqecontam.html}
that affects most the sensitivity at low energies and has been building up in the ACIS optical path
since launch in 2000 and has been developing more quickly since late 2009.
These considerations are less important in Mk~34 than in objects with softer spectra.

Independent of count rate or phase, the shapes of all 54 spectra were 
very similar in shape. For each $0.5-8$~keV spectrum, ACIS~Extract calculates the mean observed energy, $\langle\mr{E}\rangle$,
of its constituent events.
The set of these values has a narrow distribution characterised by a mean absolute deviation of 0.042~keV about a median of 
2.238~keV. The brightest and second brightest observations had $\langle\mr{E}\rangle=2.284$~keV and $\langle\mr{E}\rangle=2.305$~keV,
respectively, comfortably within the overall distribution.
Only the faintest spectra immediately after maximum
were significantly harder probably due to increased photoelectric absorption, as discussed below.

Spectral models were fit with XSPEC v12.6.0v simultaneously to all 54 spectra consisting of a constant empirical emission spectrum
modified by 54 time-dependent values of luminosity, $L_X$, and absorbing column density, $N_X$, of LMC abundances.
The emission was modelled with a
2-temperature thermal plasma of variable abundances with the best-fit parameters shown in Table~\ref{Table:XspecModel}.
The purpose of the model is not to imply the simultaneous presence of two
equilibrium plasmas of distinct temperature or to study the abundances
but instead accurately to reproduce on a phenomenological basis the shape of the underlying spectrum
in order to assess the evolution of luminosity and absorption.

\begin{figure}
\includegraphics[width=\columnwidth]{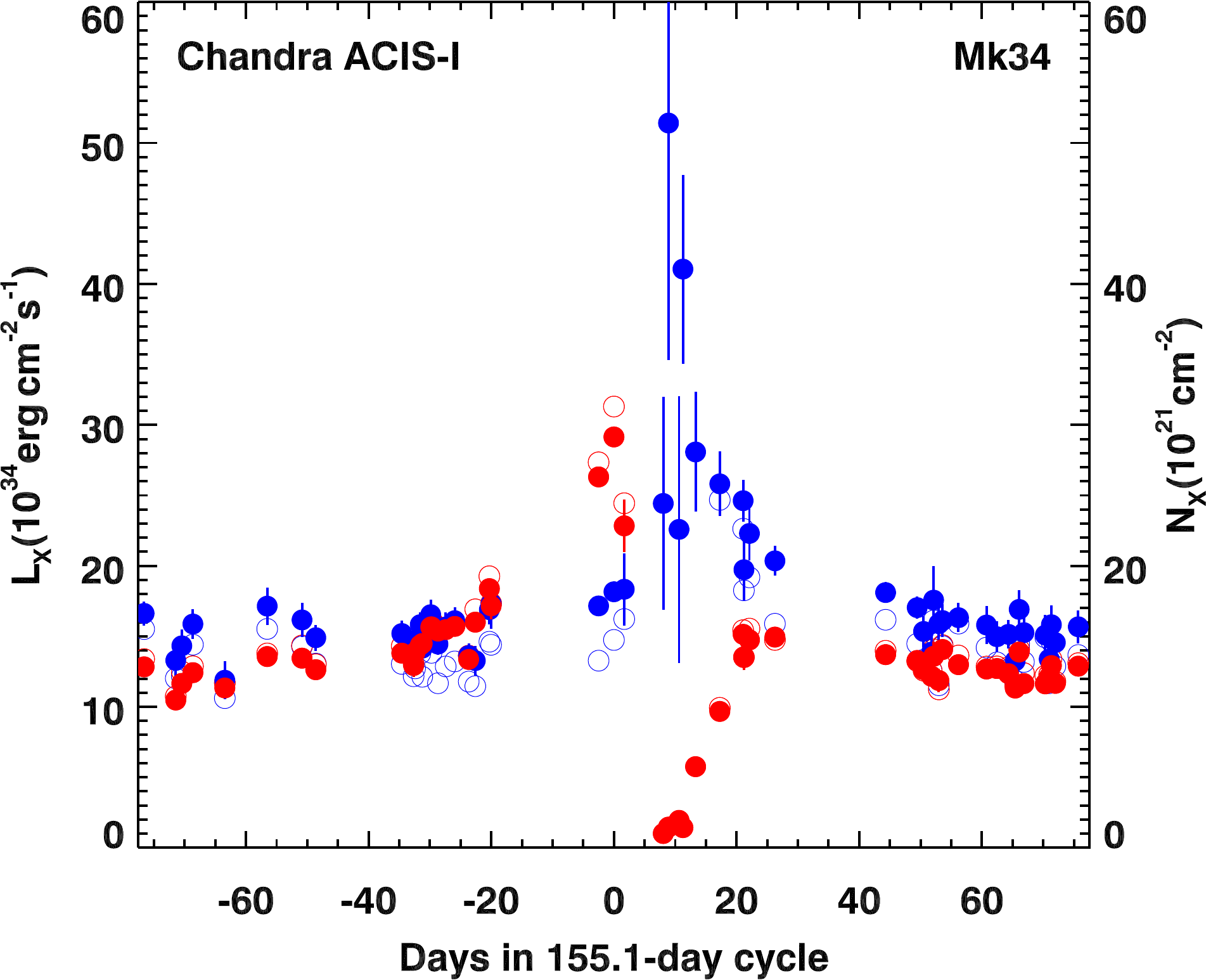}
\caption{Phase-dependent \Chandra~ACIS-I estimates of the X-ray luminosity, $L_X$, in red and absorption, $N_X$, in blue of Mk~34 in units characteristic
of colliding-wind systems in the Milky Way. The filled symbols with error bars show estimates ignoring pile-up; the open symbols without error bars were derived
from spectra reconstructed to remove the effects of pile-up.}
\label{Fig:LxNx}
\end{figure}

The 54 values of $L_X$ and $N_X$ are plotted in Figure~\ref{Fig:LxNx} in units characteristic of familiar colliding-wind binary systems
in the Galaxy of $10^{34}\mr{erg}~\mr{s}^{-1}$ and $10^{21}\mr{cm}^{-2}$ for luminosity and column density respectively.
In Mk~34, both show coherent repeatable behaviour as a function of phase.
Figure~\ref{Fig:LxNx} also suggests how pile-up has distorted the observed spectra by plotting with open symbols the values of $L_X$ and $N_X$ 
derived instead from analysis of the set of pile-up restored spectra supplied by ACIS Extract.
Recalling that pile-up affects typically 8\% and up to 17\% of counts by discarding or
moving them to higher energies, apparent hardening of the spectrum
of Mk~34 at its brightest which might have been thought due to increasing absorption was more likely caused by pile-up.

Mk~34 is an order of magnitude more luminous than any similar galactic system.
After apastron, its luminosity increases slowly at first before accelerating to reach a maximum brighter by a factor of about 3
that immediately precedes an event that looks like an eclipse. Within the limits of the data available,
the steady recovery is reproducible between cycles and shows a sharp transition at $+20$ days after periastron
to the gentle decrease that leads again to apastron.

The lack of colour variations throughout most of the orbit suggests that a constant
interstellar component might account for most of the absorption observed in most of the spectra with
a value of about $15\times10^{21}\mr{cm}^{-2}$ according to Figure~\ref{Fig:LxNx}.
This is indeed consistent with expectations from Mk34's optical and IR photometry which shows some of the highest
reddening among the hot stars in 30 Doradus.
For the narrow-band colour excess $E_{b-v}$, \citet{DCdK+:2013} reported a value of $0.47$ compared to $0.75$ from the models of \cite{HRT+:2014}.
For the elevated gas-to-dust ratio in R136 cited by \citet{DCdK+:2013} and the low metallicity of the LMC, $\mr{Z}=0.5$, also used in
the X-ray absorption model, this would imply an interstellar hydrogen
column density of $11$ or $17\times10^{21}\mr{cm}^{-2}$ for the competing values of $E_{b-v}$, bracketing the X-ray value.

Although through most of the orbit there is little evidence of X-ray colour changes,
absorption does reach a maximum of about double the interstellar value
for 2 or 3 weeks centred about 10 days after X-ray maximum
during the single X-ray eclipse revealed by the light curve when one of the stars is probably passing across the line-of-sight.
Here low count rates make some estimates uncertain
although there are some more precise measurements after eclipse egress.
Another task of future observations with more complete phase coverage will be to try to identify a second eclipse.

\input{XSPECModel}

\section{Form of the X-ray light curve}

\begin{figure}
\includegraphics[width=\columnwidth]{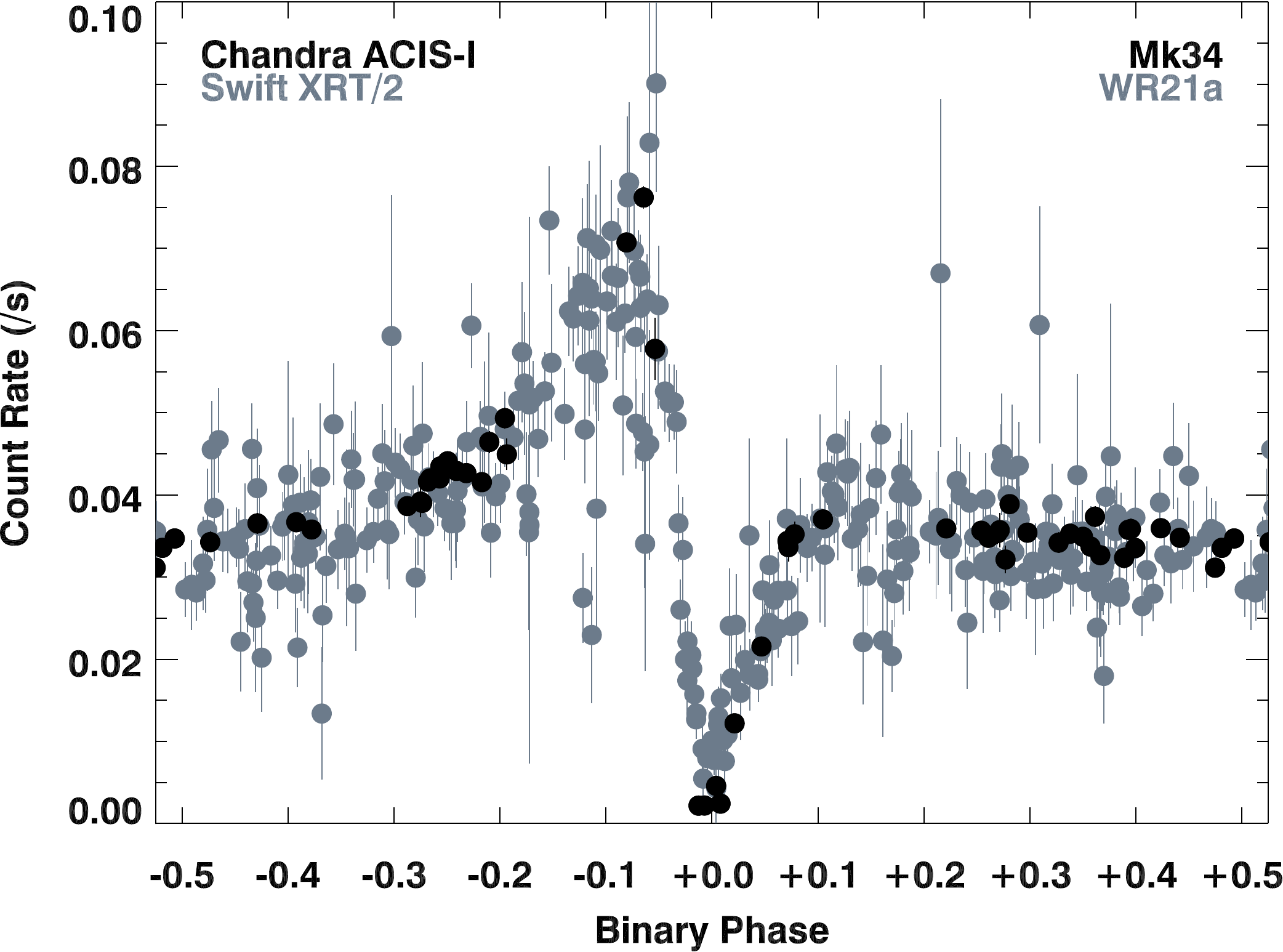}
\caption{Phase-dependent \Chandra~ACIS-I X-ray count rate of Mk~34 in comparison with the \Swift~XRT light curve of
the Galactic Wolf-Rayet eccentric binary system WR~21a after setting zero phase of Mk~34 10 days after its X-ray maximum.}
\label{Fig:Mk34WR21a}
\end{figure}

The distinctive shape of the folded X-ray light curve of Mk~34 is very similar in phased form to that seen recently in the
Galactic very massive Wolf-Rayet colliding-wind binary system WR~21a \citep{STM+:2015,GN:2016} with a combination of
100~ks of
\XMM~data at four phases and 306~ks of a \Swift~ToO XRT campaign of 330 snapshots spread evenly
over the entire 31.672-day period
of its optical radial velocity orbit \citep{NGB+:2008,TSF+:2016}.
A comparison of the two stars is shown in Figure~\ref{Fig:Mk34WR21a} where Mk~34's zero phase was shifted forward
10 days from its observed maximum count rate and the \Swift~XRT count rate was halved.
Despite the differences in period of a factor of 5 and luminosity of more than an order of magnitude,
the similarities are striking in the gradual rise to maximum followed by the subsequent deep minimum and asymmetric recovery.
The well-established spectral types of WR~21a, O3/WN6ha+O3Vz((f*)), and its Keplerian orbital elements \citep{TSF+:2016}
allow an assessment of the relationship between X-ray
orbital light-curve morphology and stellar and orbital geometry as discussed in part by \cite{GN:2016}.
Its minimum is probably caused, qualitatively at least, by some combination of three mechanisms:
absorption by the extended wind of the Wolf-Rayet star;
eclipse by its stellar core;
and reduced upstream shock velocities.
The potential utility of X-ray measurements is emphasised by the lack of eclipses at longer wavelengths
although quantitative models remain to be devised.

Once Mk~34's orbit is known in the near future, such quantitative models would seem to be required as
simple arguments, although hard to fault, appear to fail:
if the similarity of their X-ray light curves were to suggest similar orbital eccentricities for Mk~34 and WR~21a,
then, as demanded by Kepler's laws, scaling the sum of WR~21a's minimum masses of $102 \pm 6~\mr{M}_{\sun}$ \citep{TSF+:2016}
by the period ratio and the cube of the relative velocity amplitudes according to \citet{CSC+:2011}
would suggest unfeasibly
high combined minimum masses of over $1000~\mr{M}_{\sun}$ for Mk~34.

\begin{figure}
\includegraphics[width=\columnwidth]{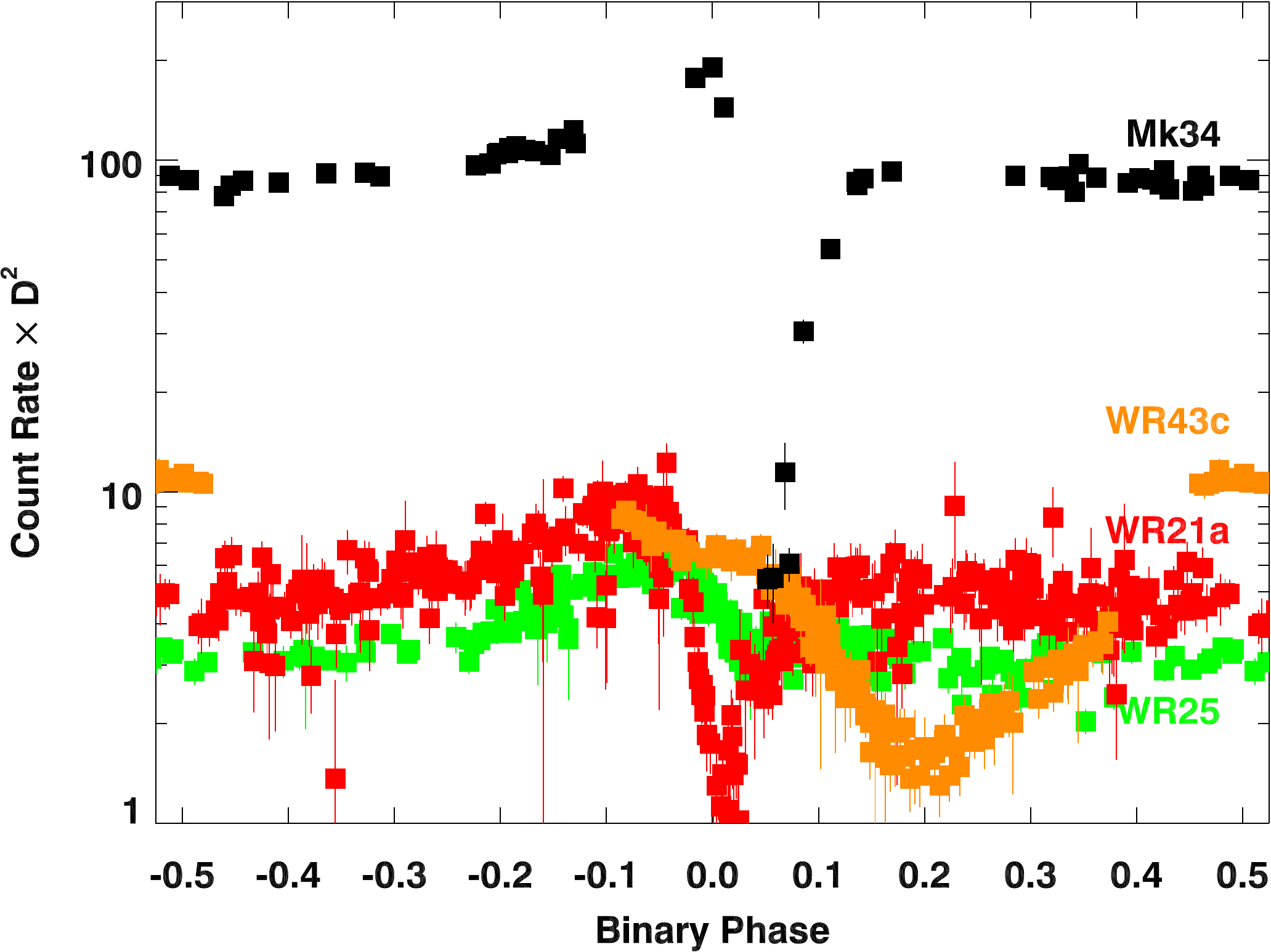}
\caption{Phase-dependent \changed{ACIS-I X-ray count rate multiplied by the square of the distance in kpc} of Mk~34 
in comparison with three other X-ray bright binary systems in the Galaxy with very massive Wolf-Rayet primary stars.}
\label{Fig:vmWRX}
\end{figure}

Mk~34 and WR~21a are not the only binary systems with very massive Wolf-Rayet primary stars that show
high-amplitude orbital phase-related X-ray variability of apparently similar type.
Of \citeauthor{CW:2011}'s (\citeyear{CW:2011}) very massive stars, WR~25 in Carina and WR~43c in
NGC~3603 are two other clear but more complex examples. The binary properties of the four systems,
with periods ranging from about 10 to 200 days, are shown in Table~\ref{Table:vmWRX}
which also includes rough estimates of the X-ray luminosities at apastron, when the stars are furthest apart;
at X-ray maximum, often close to periastron; and at eclipse minimum.
Phased X-ray light curves are plotted in Figure~\ref{Fig:vmWRX} scaled by distance to show their relative luminosities.
\Swift~XRT data of WR~21a and WR~25 were scaled by a factor of 4.0 calculated from comparison of the ACIS-I
count rate during the observation of WR~21a with ObsID 9113 with similar phases of the XRT light curve.

Mk~34 is the most luminous by about an order of magnitude, probably signalling the presence of 
two extreme stars in close proximity with high mass-loss rates and high-velocity winds.
Repeatable events with all the appearance of eclipses occur in all four systems.
For the well-established orbits of WR~21a and WR~25 these occur very close to
inferior conjunction when the primary Wolf-Rayet star is passing in front of its binary companion and presumably also, therefore,
in front of the X-ray source between the two stars.
Like Mk~34, both these stars also show increased X-ray absorption in narrow intervals around these phases.
These do not apply to the system of shortest period, WR~43c \citep{SCC+:2008},
one of the central stars in the cluster NGC~3603, 
where the smoother minimum rather occurs near a quadrature.
As argued above, once orbital geometry has been defined by Kepler's laws, X-ray eclipses by star and wind
promise direct estimates of fundamental parameters such as orbital inclination,
stellar radius and mass-loss rate such as attempted, for example, by \cite{P:2012} for WR~25.
The data shown here for WR~25 and WR~43c will be discussed in detail elsewhere.

Luminous colliding-wind X-ray sources of high-amplitude orbital-driven variability are by no means ubiquitous among
very massive Wolf-Rayet binary systems. The close binary WR~43a (WN6ha+WN6ha, P=3.7724d) in NGC3603 \citep{SCC+:2008} is normally less luminous in X-rays
than its neighbour WR~43c. Also fainter by an order of magnitude or more are the short-period WR~20a (O3If*/WN6+O3If*/WN6, P=3.686d) \citep{NRM:2008},
and the longer period WR~22 (WN7h+O9III-V, P=80.336~d) \citep{GNS+:2009} although the phase coverage of these last two systems has been poor.

Even among colliding-wind binaries in general, Mk~34 is arguably
the most luminous system yet identified. Also shown in Table~\ref{Table:vmWRX}
are comparisons with the brightest and best observed objects of this class, namely
$\eta$~Carinae \citep{HCR+:2014,CHL+:2015,CLM+:2017}, 
WR~140 \citep{PCSW:2005,SMT+:2015} and
WR~48a \citep{WvdHvW+:2012,ZTG+:2014}.

Despite much longer orbital periods of years and decades,
all show X-ray cycles with many properties in common with Mk~34 with differences only of detail.
In addition to a rich set of lines, spectra feature hard, bright X-ray continuum emission with equivalent temperatures of 4-5 keV.
After minimum at apastron, luminosities rise gradually to maximum shortly before periastron
before sudden, complex eclipses very close to periastron precede notably slow and asymmetric recoveries
towards apastron and the beginnings of a new, usually repeatable cycle.

The only rival to Mk~34's status as the most X-ray luminous of the colliding-wind binaries is the LBV system $\eta$~Carinae
which displays several characteristics that are so far
unique \citep{CHL+:2015} including flares and eclipses of variable shape. 
It was during a flare in the approach to the most recent periastron passage in 2014 that $\eta$~Carinae reached
the exceptional maximum luminosity reported in Table~\ref{Table:vmWRX}
that was about 50\% higher than in previous cycles \citep{CLM+:2017}.
For Mk~34, while the good agreement between the contrasts in count rate observed in comparison with \XMM~and \Swift~
tend to suggest a reproducible maximum, data at maximum are sparse so that
this should be subject to test with more \Chandra~data of high spatial resolution.

While Mk~34 is likely to be a system of two hydrogen-rich Wolf-Rayet stars,
the high luminosity of $\eta$~Carinae is supposed to be powered by interactions of a slow, dense LBV wind and the much faster wind
of an unidentified companion.
Nevertheless, given the obvious similarity of spectra and light curves, Mk~34 and $\eta$~Carinae probably share many aspects of shock physics and geometry.

\input{vmWRX}

\section{Conclusions}

Mk~34 is one of the most prominent Wolf-Rayet stars in the LMC and the brightest stellar X-ray source.
\Chandra~ACIS-I observations made as part of the 2~Ms T-ReX campaign on the dense stellar cluster
30 Doradus, have revealed that it has a repeatable X-ray cycle of 155.1 days that is
confirmed by archived \XMM~data and new ToO observations with \Swift.
It is the most X-ray luminous colliding-wind binary system yet identified,
exceeding even $\eta$~Carinae.

Though lacking coverage at some crucial phases,
the form of the phased X-ray light curve of Mk~34 appears repeatable and
very similar to the Galactic colliding-wind
binary system WR~21a which is of shorter period and lower luminosity but whose more detailed light curve
suggests that a combination of binary, geometrical and radiative mechanisms is responsible for the distinctive gradual rise
to maximum before the sudden onset of a deep minimum and gradual recovery.

We are in the process of establishing the
optical radial velocity orbit of Mk~34 in order to define how the geometrical disposition of the stars is related to its X-ray behaviour.
Although located 50~kpc away in the LMC, Mk~34 is a brighter X-ray source most of the time than many well-known
Wolf-Rayet systems in the Galaxy. As a result, high-resolution observations with the \Chandra~HETG will also be also feasible to
confirm a variety of low metal abundances in the LMC and study the physics and dynamics of the shocks responsible for its X-rays.
Detailed future X-ray photometry holds the promise of direct
estimates of the radius of one or both of the stellar components of Mk~34, widely thought to be among the most massive of stars.

\section*{Acknowledgements}

Broos and Townsley were supported by {\em Chandra X-ray Observatory} general observer (GO) grants GO5-6080X and GO4-15131X (PI:  L.\ Townsley), and by the Penn State ACIS Instrument Team Contract SV4-74018, issued by the {\em Chandra} X-ray Center, which is operated by the Smithsonian Astrophysical Observatory for and on behalf of NASA under contract NAS8-03060.
Partial financial support for Pollock and Tehrani was provided by the United Kingdom STFC.
We are very grateful to the \Swift~ToO program, its Project Scientist and Observatory Duty Scientists for awards
of observing time, general support and extensive use of UKSSDC data analysis tools.
\changed{
This research has made use of SAOImage DS9, developed 
by Smithsonian Astrophysical Observatory.
}





\bibliographystyle{mnras}







\bsp	
\label{lastpage}
\end{document}

%% file: Chandra.log.tex
\begin{table*}
\centering
\caption{Log of \Chandra~observations in the T-ReX survey of 30~Doradus with
observation ID and epoch;
exposure time, T;
sensitivity factor relative to the ObsID 7263 maximum;
sensitivity-corrected count rate per 1000s of Mk~34;
Kolmogorov-Smirnov variability test log-probability; and
phase interval covered, $\phi$, of the 155.1-day cycle centred on the X-ray maximum of 2014-08-22.}
\label{Table:ChandraLog}
\begin{tabular}{rccrrr@{ $\pm$ }lrr@{ $\Leftrightarrow$ }r}
\hline
ObsID & date & MJD & T(s) & factor & \multicolumn{2}{c}{Mk~34 (cts/ks)} & $\log{\mr{P}_\mr{KS}}$ & \multicolumn{2}{c}{$\phi_{155.1}$(d)} \\
\hline
16192 & 2014-05-03T04:10:27 & 56780.174 & 93761 & 0.95 &   35.9 & 0.7    &  $-0.152$ &   $+43.7$ &   $+44.9$ \\
16193 & 2014-05-08T10:15:25 & 56785.427 & 75994 & 0.95 &   35.6 & 0.7    &  $-0.003$ &   $+49.0$ &   $+49.9$ \\
16612 & 2014-05-11T02:15:31 & 56788.094 & 22672 & 0.96 &   35.4 & 1.3    &  $-0.430$ &   $+51.7$ &   $+52.0$ \\
16194 & 2014-05-12T20:00:24 & 56789.834 & 31333 & 0.93 &   38.9 & 1.2    &  $-0.732$ &   $+53.4$ &   $+53.8$ \\
16615 & 2014-05-15T08:24:45 & 56792.351 & 45170 & 0.96 &   35.4 & 1.0    &  $-0.068$ &   $+55.9$ &   $+56.5$ \\
16195 & 2014-05-24T14:09:28 & 56801.590 & 44405 & 0.93 &   33.7 & 0.9    &  $-0.040$ &   $+65.1$ &   $+65.7$ \\
16196 & 2014-05-30T00:05:56 & 56807.004 & 67109 & 0.96 &   35.8 & 0.8    &  $-0.296$ &   $+70.6$ &   $+71.4$ \\
16617 & 2014-05-31T01:27:04 & 56808.060 & 58860 & 0.93 &   33.5 & 0.8    &  $-0.255$ &   $+71.6$ &   $+72.3$ \\
16616 & 2014-06-03T22:26:17 & 56811.935 & 34530 & 0.93 &   35.9 & 1.1    &  $-0.205$ &   $+75.5$ &   $+75.9$ \\
16197 & 2014-06-06T12:32:26 & 56814.523 & 67790 & 0.95 &   34.8 & 0.8    &  $-0.292$ &   $-77.0$ &    $-76.2$ \\
16198 & 2014-06-11T20:20:49 & 56819.848 & 39465 & 0.93 &   31.1 & 1.0    &  $-0.393$ &   $-71.7$ &    $-71.2$ \\
16621 & 2014-06-14T14:46:41 & 56822.616 & 44400 & 0.93 &   34.7 & 0.9    &  $-0.203$ &   $-68.9$ &    $-68.4$ \\
16200 & 2014-06-26T20:01:47 & 56834.835 & 27361 & 0.94 &   36.5 & 1.3    &  $-0.078$ &   $-56.7$ &    $-56.4$ \\
16201 & 2014-07-21T22:13:45 & 56859.926 & 58390 & 0.93 &   42.0 & 0.9    &  $-0.171$ &   $-31.6$ &    $-30.9$ \\
16640 & 2014-07-24T11:21:26 & 56862.473 & 61679 & 0.93 &   44.1 & 0.9    &  $-0.018$ &   $-29.1$ &    $-28.3$ \\
16202 & 2014-08-19T15:30:01 & 56888.646 & 65128 & 0.93 &   70.7 & 1.1    &  $-0.313$ &     $-2.9$ &      $-2.1$ \\
17312 & 2014-08-22T06:21:18 & 56891.265 & 44895 & 0.93 &   76.2 & 1.4    &  $-0.346$ &     $-0.3$ &     $+0.3$ \\
16203 & 2014-09-02T12:47:11 & 56902.533 & 41423 & 0.94 &     2.4 & 0.3    &  $-0.014$ &   $+11.0$ &   $+11.5$ \\
17413 & 2014-09-08T15:21:28 & 56908.640 & 24650 & 0.94 &   21.6 & 1.0    &  $-0.151$ &  $+17.1$ &   $+17.4$ \\
17414 & 2014-09-13T12:24:59 & 56913.517 & 17317 & 0.94 &   35.2 & 1.5    &  $-0.078$ &  $+22.0$ &   $+22.2$ \\
16442 & 2014-10-25T13:38:44 & 56955.569 & 48350 & 0.93 &   34.9 & 0.9    &  $-1.161$ &  $+64.0$ &   $+64.6$ \\
17545 & 2014-10-28T04:14:57 & 56958.177 & 34530 & 0.93 &   32.7 & 1.1    &  $-0.086$ &   $+66.6$ &   $+67.1$ \\
17544 & 2014-11-01T16:52:08 & 56962.703 & 25642 & 0.93 &   35.9 & 1.3    &  $-0.326$ &   $+71.2$ &   $+71.5$ \\
16443 & 2014-11-14T23:14:31 & 56975.968 & 34530 & 0.93 &   33.6 & 1.1    &  $-0.322$ &   $-70.7$ &   $-70.2$ \\
17486 & 2014-12-04T13:39:50 & 56995.569 & 33541 & 0.92 &   36.7 & 1.1    &  $-0.353$ &   $-51.1$ &   $-50.6$ \\
17555 & 2014-12-06T16:40:37 & 56997.695 & 55247 & 0.90 &   35.8 & 0.9    &  $-0.017$ &   $-48.9$ &   $-48.3$ \\
17561 & 2014-12-20T17:22:40 & 57011.724 & 54567 & 0.93 &   38.6 & 0.9    &  $-0.154$ &   $-34.9$ &   $-34.3$ \\
17562 & 2014-12-25T15:11:01 & 57016.633 & 42031 & 0.92 &   43.5 & 1.1    &  $-0.293$ &   $-30.0$ &   $-29.5$ \\
16444 & 2014-12-27T22:58:58 & 57018.958 & 41440 & 0.88 &   42.9 & 1.1    &  $-1.282$ &   $-27.7$ &   $-27.2$ \\
16448 & 2015-02-14T11:54:08 & 57067.496 & 34599 & 0.95 &   34.4 & 1.1    &  $-1.223$ &   $+20.9$ &   $+21.3$ \\
17602 & 2015-02-19T13:57:46 & 57072.582 & 51705 & 0.95 &   37.0 & 0.9    &  $-0.130$ &   $+25.9$ &   $+26.6$ \\
16447 & 2015-03-26T05:26:59 & 57107.227 & 26868 & 0.94 &   34.2 & 1.3    &  $-1.138$ &   $+60.6$ &   $+60.9$ \\
16199 & 2015-03-27T20:27:05 & 57108.852 & 39461 & 0.94 &   35.3 & 1.0    &  $-0.853$ &   $+62.2$ &   $+62.7$ \\
17640 & 2015-03-31T13:14:43 & 57112.552 & 26318 & 0.92 &   37.3 & 1.3    &  $-0.807$ &   $+65.9$ &   $+66.2$ \\
17641 & 2015-04-04T19:45:40 & 57116.823 & 24638 & 0.95 &   32.3 & 1.2    &  $-0.456$ &   $+70.2$ &   $+70.5$ \\
16445 & 2015-05-27T00:18:12 & 57169.013 & 49310 & 0.95 &   39.0 & 0.9    &  $-0.412$ &   $-32.7$ &   $-32.1$ \\
17660 & 2015-05-29T14:55:28 & 57171.622 & 38956 & 0.95 &   42.0 & 1.1    &  $-0.202$ &   $-30.1$ &   $-29.6$ \\
16446 & 2015-06-02T11:50:14 & 57175.493 & 47547 & 0.95 &   42.6 & 1.0    &  $-0.316$ &   $-26.2$ &   $-25.7$ \\
17642 & 2015-06-08T05:11:14 & 57181.216 & 34438 & 0.92 &   49.3 & 1.3    &  $-0.584$ &   $-20.5$ &   $-20.1$ \\
16449 & 2015-09-28T05:35:14 & 57293.233 & 24628 & 0.92 &   34.3 & 1.3    &  $-0.127$ &   $-63.6$ &   $-63.3$ \\
18672 & 2015-11-08T01:04:22 & 57334.045 & 30574 & 0.92 &   46.4 & 1.3    &  $-0.291$ &   $-22.8$ &   $-22.4$ \\
18706 & 2015-11-10T17:09:59 & 57336.715 & 14776 & 0.92 &   44.9 & 1.9    &  $-0.068$ &   $-20.1$ &   $-19.9$ \\
18720 & 2015-12-02T10:49:02 & 57358.451 &   9832 & 0.51 &   57.8 & 3.8    &  $-0.922$ &      $+1.6$ &    $+1.7$ \\
18721 & 2015-12-08T17:13:14 & 57364.718 & 25598 & 0.74 &     2.2 & 0.4    &  $-0.067$ &      $+7.9$ &    $+8.2$ \\
17603 & 2015-12-09T15:27:36 & 57365.644 & 13778 & 0.71 &     2.2 & 0.6    &  $-0.332$ &      $+8.8$ &    $+9.0$ \\
18722 & 2015-12-11T09:09:39 & 57367.382 &   9826 & 0.62 &     4.6 & 1.1    &  $-0.739$ &     $+10.5$ &   $+10.7$ \\
18671 & 2015-12-13T23:41:12 & 57369.987 & 25617 & 0.55 &   12.2 & 1.0    &  $-0.234$ &   $+13.1$ &   $+13.5$ \\
18729 & 2015-12-21T22:10:30 & 57377.924 & 16742 & 0.74 &   33.6 & 1.7    &  $-2.099$ &   $+21.1$ &   $+21.3$ \\
18750 & 2016-01-20T00:41:30 & 57407.029 & 48318 & 0.85 &   34.8 & 0.9    &  $-5.557$ &   $+50.2$ &   $+50.8$ \\
18670 & 2016-01-21T20:59:37 & 57408.875 & 14565 & 0.63 &   35.7 & 2.1    &  $-0.758$ &   $+52.0$ &   $+52.2$ \\
18749 & 2016-01-22T16:14:19 & 57409.677 & 22153 & 0.56 &   32.1 & 1.7    &  $-0.423$ &   $+52.8$ &   $+53.1$ \\
\hline
 5906 & 2006-01-21T19:04:02 & 53756.794 & 12317 & 1.00 &   39.2 & 1.9    &  $-0.992$ &   $-32.7$ &   $-32.6$ \\
 7263 & 2006-01-22T16:51:51 & 53757.703 & 42528 & 1.00 &   41.6 & 1.0    &  $-0.169$ &   $-31.8$ &   $-31.3$ \\
 7264 & 2006-01-30T15:06:27 & 53765.629 & 37593 & 0.98 &   41.5 & 1.1    &  $-0.885$ &   $-23.9$ &   $-23.5$ \\
 \hline
\end{tabular}
\end{table*}

%% file: XMM.log.tex
\begin{table*}
\centering
\caption{Log of \XMM~observations of Mk~34 with
observation ID and epoch;
exposure time, T;
detected XSA PPS count rate per 1000s;
phase interval covered, $\phi$, of the 155.1-day cycle;
and ACIS-I count rate from the nearest \Chandra~observation in phase.}
\label{Table:XMMLog}
\begin{tabular}{ccccrr@{ $\pm$ }lr@{ $\pm$ }lr@{ $\pm$ }lr@{ $\Leftrightarrow$ }rr@{ $\pm$ }l}
\hline
rev & ObsID & date & MJD & T(s) & \multicolumn{2}{c}{EPIC-pn} 
                                                      & \multicolumn{2}{c}{EPIC-MOS1} 
                                                      & \multicolumn{2}{c}{EPIC-MOS2} 
                                                      & \multicolumn{2}{c}{$\phi_{155.1}$(d)}
                                                      & \multicolumn{2}{c}{ACIS-I}  \\
      &  & & &  & \multicolumn{2}{c}{(cts/ks)} 
                                                      & \multicolumn{2}{c}{(cts/ks)} 
                                                      & \multicolumn{2}{c}{(cts/ks)} 
                                                      & \multicolumn{2}{c}{} 
                                                      & \multicolumn{2}{c}{(cts/ks)} \\
\hline
0357  & 0113020201 & 2001-11-19 & 52232.9 & 38159 &   \NaN{2}   &  124.4 & 2.5 & 125.0 & 2.5 &   $-5.6$ &   $-5.2$ & 70.7 & 1.1 \\
2138  & 0679380101 & 2011-08-13 & 55786.3 & 24268 &    275.5 & 18.5   &  \NaN{2} & \NaN{2} &   $-19.5$ &   $-19.3$ & 44.9 & 1.9 \\
2358  & 0690744401 & 2012-10-23 & 56223.8 & 38159 &   192.2 & 4.7    &    64.8 & 2.7 &   66.4 & 2.7 &   $-47.4$ &   $-46.9$ & 35.8 & 0.9 \\
 \hline
\end{tabular}
\end{table*}

%% file: Swift.log.tex
\begin{table*}
\centering
\caption{Log of \Swift~ToO XRT observations of Mk~34 near the predicated maximum in 2016 May with
observation ID and epoch;
exposure time, T;
detected XRT count rate per 1000s; and
phase interval covered, $\phi$, of the 155.1-day cycle.}
\label{Table:SwiftLog}
\begin{tabular}{lcccrr@{ $\pm$ }lr@{ $\Leftrightarrow$ }r}
\hline
target & ObsID & date & MJD & T(s) & \multicolumn{2}{c}{XRT} 
                                            & \multicolumn{2}{c}{$\phi_{155.1}$(d)} \\
     &      & & &  & \multicolumn{2}{c}{(cts/ks)} 
                                                     & \multicolumn{2}{c}{}  \\
\hline
Mk~34          & 00034533002 & 2016-05-03 & 57511.3 &  7085 &  38.8 & 2.9 &       $-0.63$ &     $+0.06$ \\
Mk~34          & 00034533003 & 2016-05-10 & 57518.1 &  2566 &    6.0 & 2.4 &      $+6.15$ &     $+6.95$ \\
SNR N157B  & 00081912001 & 2016-05-10 & 57518.2 &    955 &  10.3 & 4.0 &      $+6.28$ &     $+6.36$ \\
Mk~34          & 00034533004 & 2016-05-11 & 57519.2 &   3481 &   8.2 & 2.3 &      $+7.28$ &     $+8.02$ \\
Mk~34          & 00034533005 & 2016-05-15 & 57523.1 &   2469 &   7.4 & 2.6 &    $+11.13$ &   $+12.13$ \\
Mk~34          & 00034533006 & 2016-05-31 & 52232.9 &     787 &   9.0 & 5.0 &    $+27.71$ &   $+27.90$ \\
Mk~34          & 00034533007 & 2016-06-05 & 52232.9 &   6834 & 19.9 & 2.1 &    $+32.98$ &   $+34.01$ \\
 \hline
\end{tabular}
\end{table*}

%% file: XSPECModel.tex
\begin{table}
\centering
\caption{Best-fit \texttt{XSPEC} reference plasma emission model of Mk~34.}
\label{Table:XspecModel}
\begin{tabular}{lr@{.}l@{ $\pm$ }r@{.}l}
\hline
\multicolumn{5}{c}{$A_{1}\mathtt{vapec}({\mr kT}_1)+A_{2}\mathtt{vapec}(\mr{kT}_2)$} \\
\hline
     ${\rm kT}_1$  & 1&198& 0&040~keV \\
             $A_{1}$  & 2&251& 0&314$\times10^{-4}~\mr{cm}^{-5}$ \\
     ${\rm kT}_2$  & 4&460& 0&209~keV \\
             $A_{2}$  & 4&675& 0&219$\times10^{-4}~\mr{cm}^{-5}$ \\
                     Ne  &  0&      & 0&072   \\
                     Mg  &  0&552& 0&125   \\
                       Al  &  0&      & 0&330   \\
                       Si  &  0&756& 0&094   \\
                        S  &  0&742& 0&100   \\
                       Ar  &  0&727& 0&305   \\
                      Ca  &  0&109 & 0&402  \\
                      Fe  &  0&336& 0&029   \\
\hline
\end{tabular}
\end{table}

%% file: vmWRX.tex
\begin{table*}
\centering
\caption{Properties of Mk~34 and other colliding-wind systems.}
\label{Table:vmWRX}
\begin{tabular}{lr@{+}llr@{.}lr@{.}lr@{.}lr@{.}lr@{.}lr@{.}l}
\hline
system &  \multicolumn{2}{c}{spectral type}
          &  \multicolumn{1}{c}{cluster}
          & \multicolumn{2}{c}{distance}
          & \multicolumn{2}{c}{period} 
          & \multicolumn{2}{c}{$e$} 
           & \multicolumn{6}{c}{$\mr{L}_\mr{X}$}  \\
          & \multicolumn{2}{c}{}
          & 
          & \multicolumn{2}{c}{(kpc)}
          & \multicolumn{2}{c}{(d)} 
          & \multicolumn{2}{c}{} 
           & \multicolumn{6}{c}{($10^{34}\mr{erg}~\mr{s}^{-1}$)}   \\
          & \multicolumn{2}{c}{}
          & 
          & \multicolumn{2}{c}{}
          & \multicolumn{2}{c}{}
          & \multicolumn{2}{c}{}
          & \multicolumn{2}{c}{$\Rightarrow$}
          & \multicolumn{2}{c}{$\Uparrow$} 
           & \multicolumn{2}{c}{$\Downarrow$}   \\
 \hline
Mk~34           & WN5ha              & WN5ha      & 30~Doradus        & 50 &   & 155 & 1     & \multicolumn{2}{c}{...} &  12 & & 32 & & 1 & \\
\hline
WR~43c        & O3If*/WN6         & ?                & NGC~3603          &  7 & 6 &     8 & 89   & 0 & 30   &  0 & 9 & \multicolumn{2}{c}{...} & 0 & 1 \\
WR~21a        &  O2.5If*/WN6ha & OVz((f*))   & (Westerlund 2)   &  4 & 16 &   31 & 672 & 0 & 6949 &  0 & 6 & 1 & 2 & 0 & 1 \\
WR~25          & O2.5If*/WN6      & ?                & Trumpler~16       &  2 & 3 & 207 & 808 & 0 & 4458 &  0 & 4 & 0 & 9 & 0 & 4 \\
 \hline
$\eta$~Carinae & LBV              & (WR)         & Trumpler~16       &  2 & 3 & 2202 & 7 & 0 & 9 & 3 & 8 & 26 & & 0 & 2 \\
WR~48a    & WC8              & (O)         & G305       &  4 & 0 &  12000 & & \multicolumn{2}{c}{...}  & 1 & 9 & \multicolumn{2}{c}{...} & 0 & 1 \\
WR~140    & WC7              & O4.5       & \multicolumn{1}{c}{...}    &  1 & 67 & 2896 & 35 & 0 & 8964 & 1 & 0 & 4 & 1 & 1 & 0 \\
 \hline
 \multicolumn{15}{l}{Luminosity event signifiers: $\Rightarrow$ at apastron; $\Uparrow$ at maximum; $\Downarrow$ at eclipse minimum.} \\
\end{tabular}
\end{table*}